\documentclass[conference]{IEEEtran}

\usepackage{cite}
\usepackage{amsmath,amssymb,amsfonts}
\usepackage{algorithmic}
\usepackage{graphicx}
\usepackage{multirow}
\usepackage{textcomp}
\usepackage[inline]{enumitem} 
\usepackage{url}
\usepackage{xcolor}
\usepackage{verbatim}
\usepackage{authblk}

\def\BibTeX{{\rm B\kern-.05em{\sc i\kern-.025em b}\kern-.08em
    T\kern-.1667em\lower.7ex\hbox{E}\kern-.125emX}}
\begin{document}

\title{Exploring Local Interpretable Model-Agnostic Explanations for Speech Emotion Recognition with Distribution-Shift}

\author[2,3\IEEEauthorrefmark{1}]{Maja J. Hjuler}
\author[1]{Line H. Clemmensen}
\author[1\textdagger]{Sneha Das}
\affil[1]{Dept. of Applied Mathematics and Computer Science, Technical University of Denmark, 2800 Lyngby, Denmark}
\affil[2]{University Grenoble Alpes, CNRS, Grenoble INP, LIG, 38000 Grenoble, France } 
\affil[3]{School of Computer Science, Queensland University of Technology, Brisbane QLD 4000, Australia \authorcr Email: {\tt maja-jonck.hjuler@univ-grenoble-alpes.fr, lkhc@dtu.dk, sned@dtu.dk}} 


\maketitle
\begingroup\renewcommand\thefootnote{\IEEEauthorrefmark{1}}
\footnotetext{The author was affiliated with the Technical University of Denmark when this work was carried out.}
\endgroup
\begingroup\renewcommand\thefootnote{\textdagger}
\footnotetext{Corresponding Author}
\endgroup

\begin{abstract}
We introduce EmoLIME\footnote{Source code: https://github.com/snehadas/EmoLIME}, a version of local interpretable model-agnostic explanations (LIME) for black-box Speech Emotion Recognition (SER) models.
To the best of our knowledge, this is the first attempt to apply LIME in SER.
EmoLIME generates high-level interpretable explanations and identifies which specific frequency ranges are most influential in determining emotional states. 
The approach aids in interpreting complex, high-dimensional embeddings such as those generated by end-to-end speech models.
We evaluate EmoLIME, qualitatively, quantitatively, and statistically, across three emotional speech datasets, using classifiers trained on both hand-crafted acoustic features and Wav2Vec 2.0 embeddings. We find that EmoLIME exhibits stronger robustness across different models than across datasets with distribution shifts, highlighting its potential for more consistent explanations in SER tasks within a dataset.

\end{abstract}


\section{Introduction}

Transformer models have revolutionized large-scale signal processing, influencing all data modalities [\citen{Baevski_2020}, \citen{Hsu_2021}, \citen{Chen_2021}], including speech and audio signals [\citen{Pepino_2021}, \citen{Wagner_2023}, \citen{zhang_2021}].
While they are versatile across different domains due to their ability to incorporate information and structure in over-parameterized spaces, this also leads to black-box decisions, which is one of their main drawbacks. In other words, it is non-trivial to understand the decision-making process in transformers. 
In contrast to hand-crafted features, deep features may not represent any physical interpretation, and require alternative explainability techniques to aid the transparency and understanding behind the automated decisions.

\begin{figure}[!tbh]

\begin{minipage}[b]{1.0\linewidth}
  \centering
  \centerline{\includegraphics[width=\linewidth]{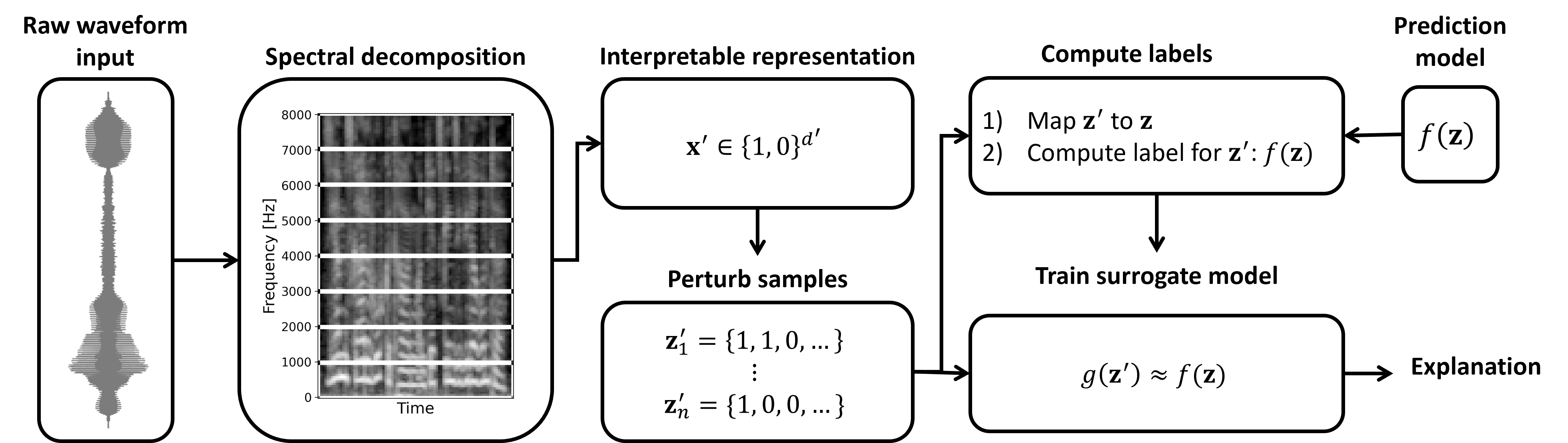}}
\end{minipage}
\caption{Functional block diagram of EmoLIME inspired by \cite{mishra2017a} and  \cite{haunschmid2020audiolime}.}
\label{fig:schematic}
%
\end{figure}

Explainable Artificial Intelligence (XAI) is rapidly advancing due to the importance of understanding the decision-making process of black-box deep-learning and machine learning models. This is particularly critical in high-stakes sectors such as healthcare, law, and education, where the {\it model outcome is as important as how one arrived at it}. Transparency and explainability of automated systems are now also necessitated through regulatory mandates and frameworks like the EU AI act and the OECD AI principles, respectively [\citen{EU_AI_Act_2024}, \citen{OECD_AI_principle}]. 

XAI techniques are widely researched and established in computer vision (CV) and Natural Language Processing (NLP). Due to the tangible and physical nature of visual and text data, defining connections between input and output through models, and thereby explaining model predictions, is relatively more intuitive. This is in contrast to speech and audio signals, where XAI methods need to consider {\it what to explain?}; this is further influenced by the corresponding speech processing task and its application. Therefore, only a few XAI methods developed for CV and NLP are directly transferable to speech processing. 

\begin{figure}[!tbh]

\begin{minipage}[b]{0.43\linewidth}
  \centering
  \centerline{\includegraphics[width=4.0cm]{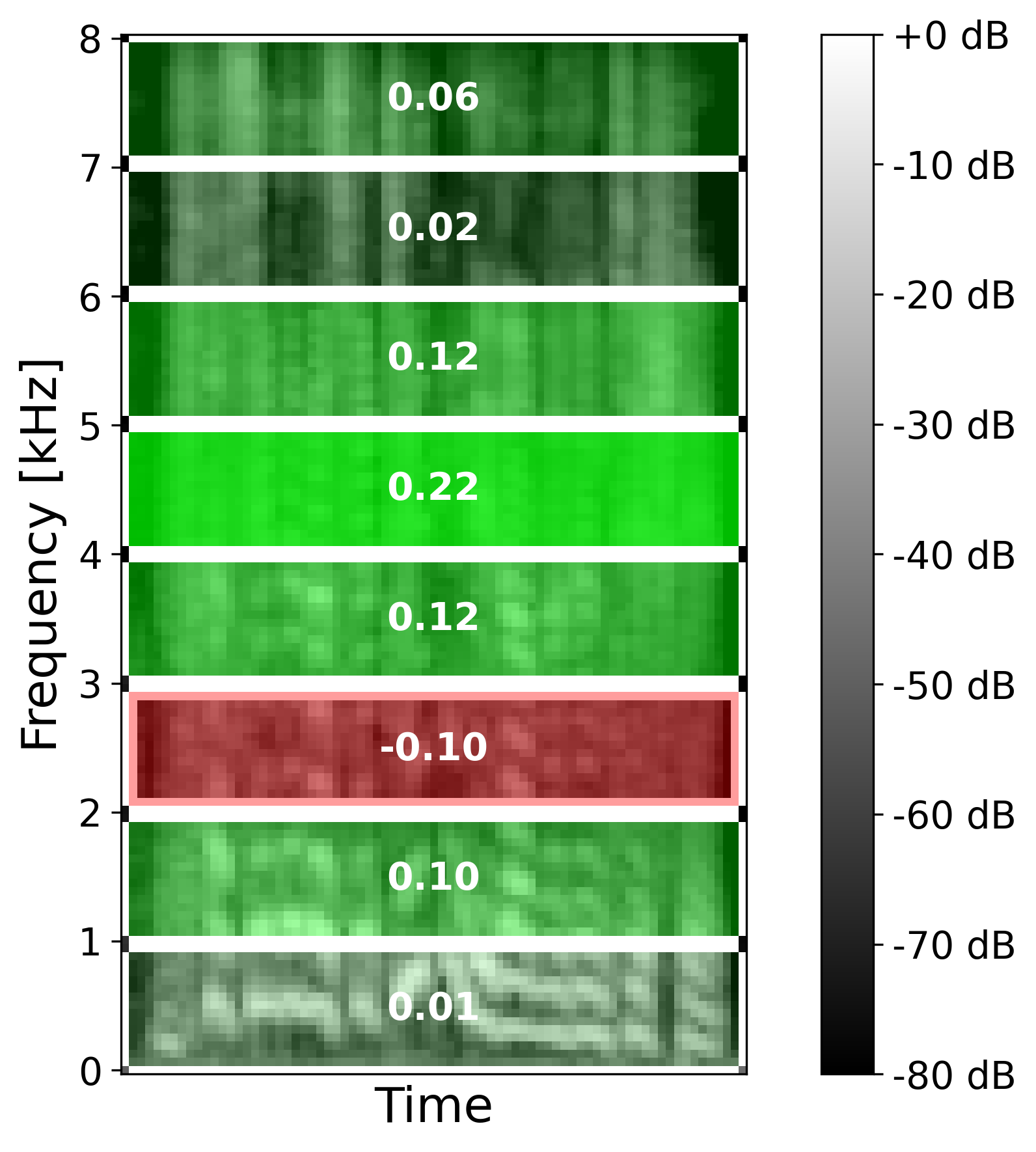}}
  \centerline{(a) \texttt{wav2vec2-SVC}}\medskip
\end{minipage}
\hfill
\begin{minipage}[b]{.43\linewidth}
  \centering
  \centerline{\includegraphics[width=4.0cm]{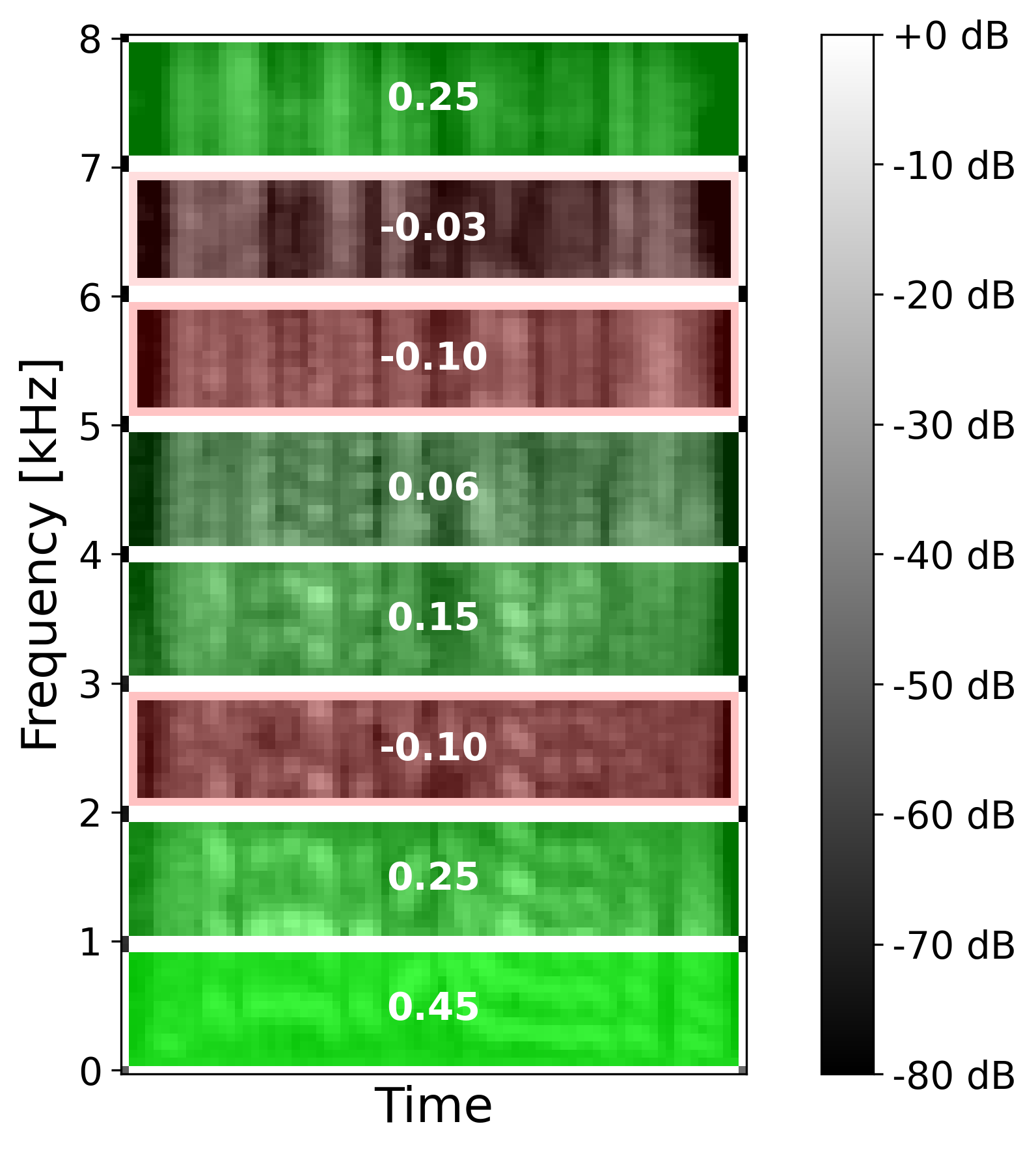}}
  \centerline{(b) \texttt{ComParE-SVC}}\medskip
\end{minipage}
\caption{Example explanations for the \textbf{happy} expression of a German sentence from EMODB. 
Components highlighted in green account for a true prediction. Weights are annotated in white.
a) Higher weight is given to high-pitch sounds (high frequency) for \texttt{wav2vec2-SVC}. 
b) The same pattern cannot be recognized for the \texttt{ComParE-SVC} model.
}
\label{fig:spectral_decomps_1}
\end{figure}

LIME (Local Interpretable Model-Agnostic Explanations) and SHAP ({SH}apley {A}dditive ex{P}lanations) are state-of-the-art XAI methods [\citen{ribeiro_2016}, \citen{lundberg2017_shap}] and are model-agnostic, ie: they can be applied to any machine-learning model. Hence, they have also been explored within speech-based classification models; LIME has been adapted to Automatic Speech Recognition (ASR) \cite{wu2024a} and 
SHAP has been employed in speech emotion recognition (SER) to evaluate feature importance [\citen{nfissi2024a}, \citen{adeel2024a}].
In contrast to gradient-based XAI techniques, LIME has an advantage in explaining waveform-fed models by directly assigning importance to decomposed audio patches rather than single time points \cite{das2022zero}.
This makes LIME explanations more aligned with human intuition and easier to interpret since we can relate different elements or segments of the audio to the prediction.
SHAP was first proposed as a uniﬁed framework for interpreting predictions and it is based on Shapley values from game theory. 
Some disadvantages of SHAP when compared to LIME include a lack of intuitiveness when working with complex transformed features from deep learning models that do not directly represent any physical characteristics of the audio.
If the end-users are non-technical experts, even hand-crafted features like Mel-frequency cepstral coeﬃcients (MFCCs) may not be considered interpretable. Furthermore, the technique can be computationally expensive for high-dimensional datasets and multi-class classification. The hand-crafted feature sets can consist of thousands of acoustic parameters making SHAP infeasible depending on system memory constraints.


In this work, we present EmoLIME, to explain the predictions of SER classifiers, developed for both hand-crafted and deep features. 
Due to the relevance of frequency based features in SER (eg: tone, pitch, etc), we primarily focus on spectral decomposition.   
 EmoLIME is developed on LIME 
 by decomposing the audio into equally sized frequency components. 
This leads to spectral masking in the training of the surrogate model. 
Explanations are generated by perturbing the input and training a linear sparse surrogate model which assigns weights to each input component. 
Our main {\it contributions} are summarised as follows:
\begin{enumerate*}
    \item We introduce EmoLIME, a LIME technique for interpretable local explanations of black-box SER models.
    To the best of our knowledge, our work represents the first attempt to apply LIME in SER.
    \item We demonstrate EmoLIME on three emotional speech datasets for classifiers trained on hand-crafted and deep features, i.e.~embeddings from a general speech model.
    \item We investigate the transferability of the explanations across three datasets, with statistical conclusions on the influence of distribution shifts on the explanations. 
\end{enumerate*}

\begin{figure}[htbp]

\begin{minipage}[b]{0.49\linewidth}
  \centering
  \centerline{\includegraphics[width=4.0cm]{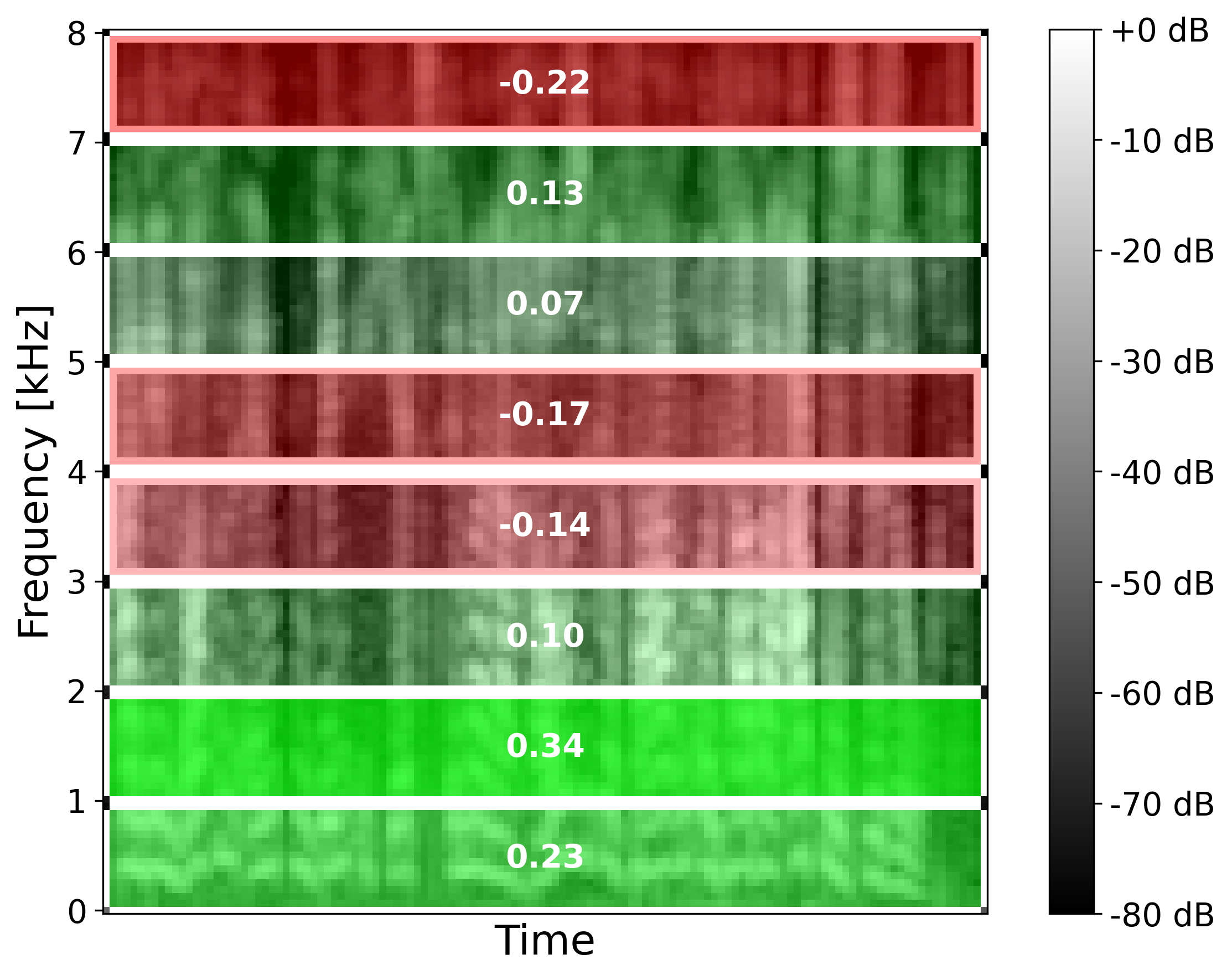}}
  \centerline{(a) \texttt{wav2vec2-SVC}}\medskip
\end{minipage}
\hfill
\begin{minipage}[b]{.49\linewidth}
  \centering
  \centerline{\includegraphics[width=4.0cm]{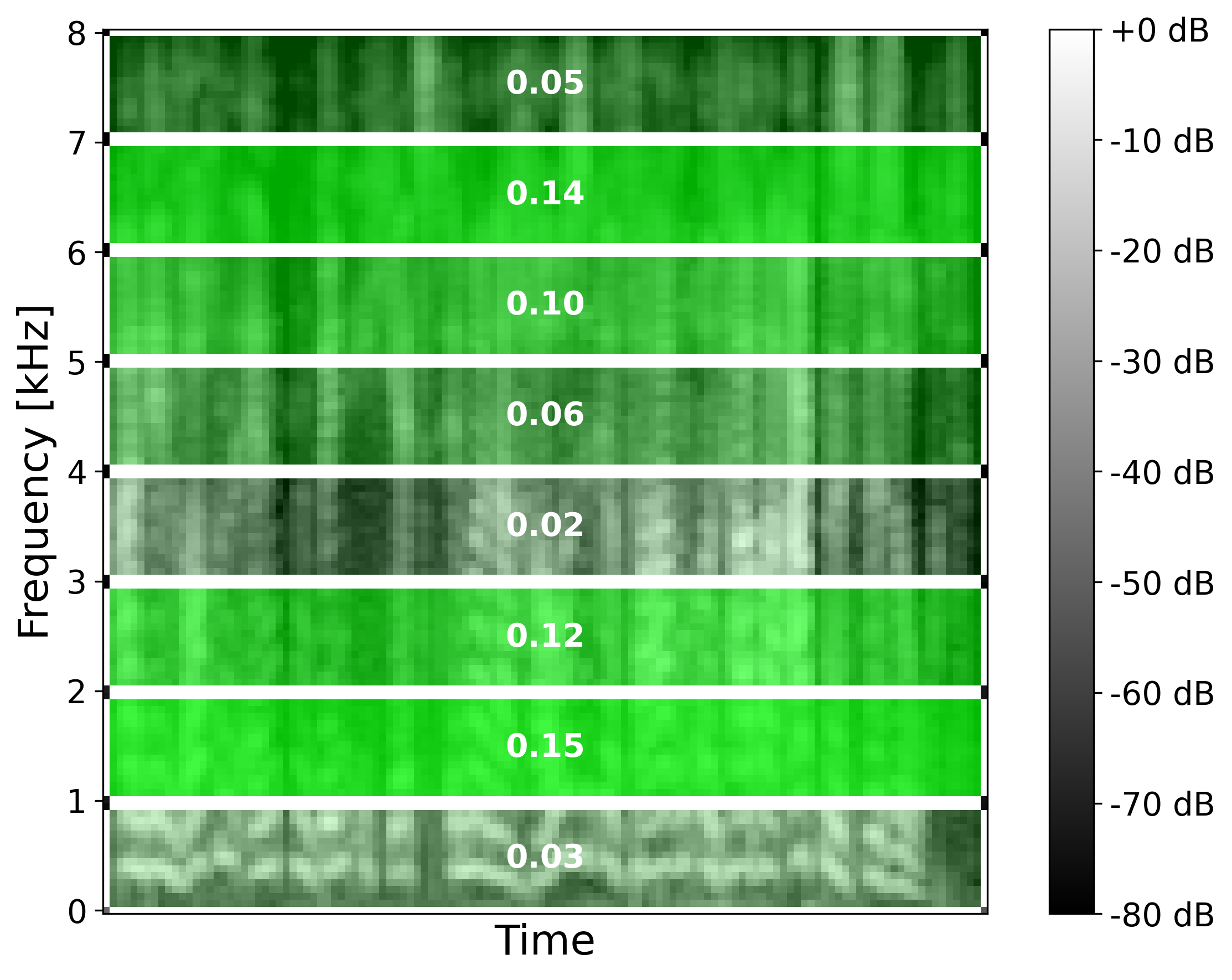}}
  \centerline{(b) \texttt{ComParE-SVC}}\medskip
\end{minipage}
\caption{Explanations for the \textbf{angry} expression of a German sentence from EMODB. 
a) More weight is given to low-pitch sounds (low frequency) for \texttt{wav2vec2-SVC}. 
b) Weights are more uniformly distributed for the \texttt{ComParE-SVC} model.
}
\label{fig:spectral_decomps_2}
\end{figure}

\section{Related Work}
XAI methods are often classified by the stage of application (before, during, or after model training), the scope (local or global), and the input data format \cite{vilone_2020}.
The explanations can also have different formats including numerical, logical, visual, and textual. 
Depending on the input audio representation (waveform, spectrogram, etc.) different XAI methods are applicable. 

In a recent review \cite{akman2024audio}, existing XAI methods for audio models are summarized and the authors emphasize the importance of enhancing their interpretability and trust.
XAI methods are split into two categories: generic XAI methods, e.g.~Integrated gradients \cite{sundararajan2017}, LIME \cite{ribeiro_2016}, and SHAP \cite{lundberg2017_shap}, and XAI methods specialized for audio models, e.g.~LRP \cite{becker2023_LRP} and DFT-LRP \cite{frommholz2023_DFT_LRP}. 
Common to methods is they aim to explain complex audio signals and leverage human adeptness at interpreting harmonies, rhythm, and other high-level concepts through listening. 
SoundLIME (SLIME) proposed in \cite{mishra2017a} extends LIME to music content analysis, specifically to singing voice detection. 
Furthermore, LIME was proposed for audio classification in AudioLIME \cite{haunschmid2020audiolime}, a system that uses source separation to produce listenable explanations.
Recently, an application of LIME to generate faithful audio explanations for COVID-19 detection from recordings of patients' coughs was presented in CoughLIME \cite{wullenweber2022coughlime}.
What sets these studies apart is the classification task that LIME is extended to and the type of segmentation applied in the algorithm. 
While AudioLIME separates the audio into different sources, SLIME and CoughLIME decompose the input data into temporal, frequency, and time-frequency segmentations.
The AudioLIME implementation does not generalize to emotional speech data from a single speaker, i.e.~a single source.
Furthermore, SLIME and CoughLIME generate explanations for binary classifiers, which are not directly applicable to multi-class SER models. 

\section{Method}

\begin{figure*}[!htb]
\centering
\begin{minipage}[b]{0.86\textwidth}
  \centering
  \centerline{\includegraphics[width=\textwidth]{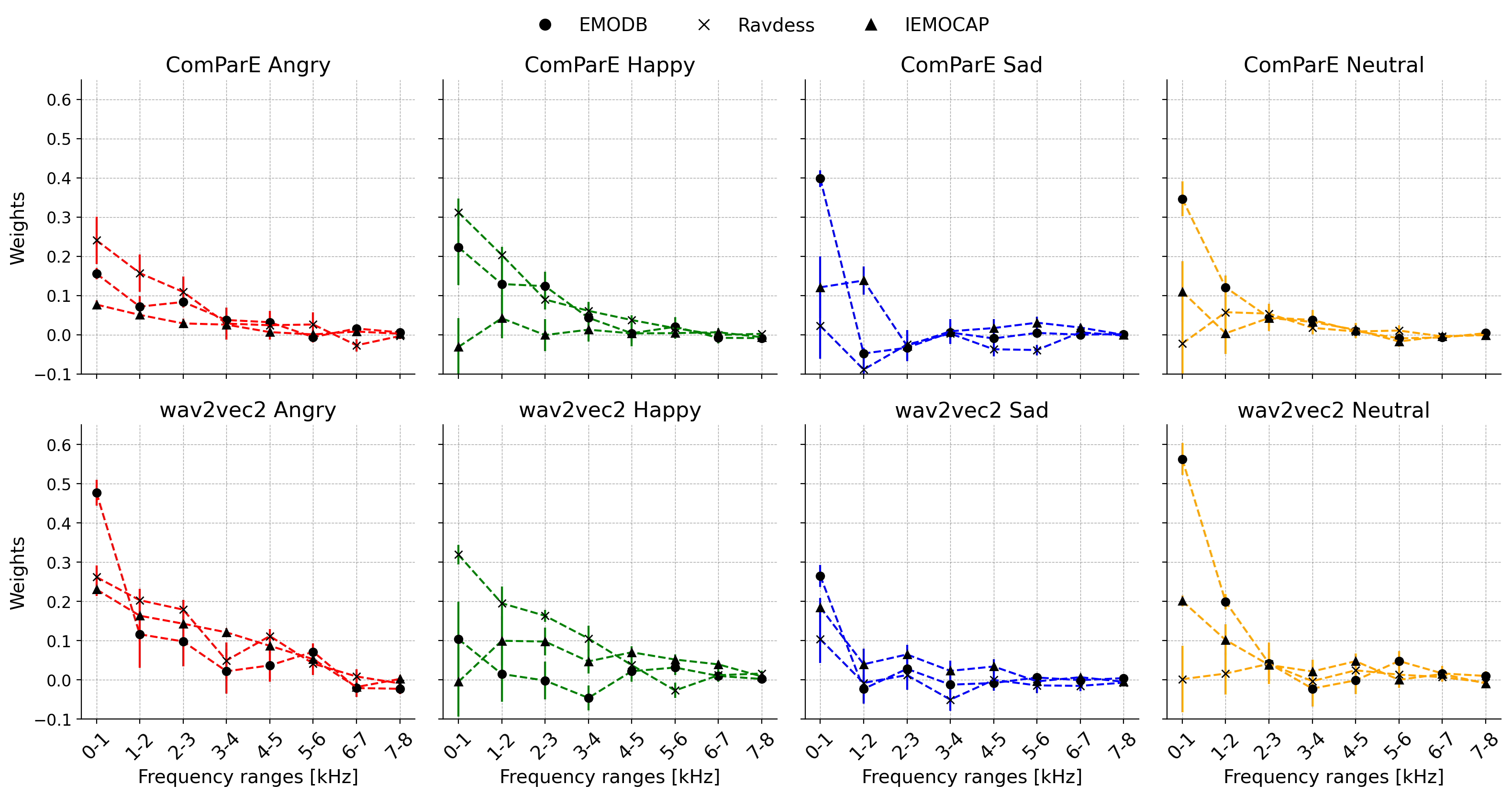}}
\end{minipage}
%
\caption{Comparison of spectral decomposition weights for the models based on ComParE (top) vs.~deep features (bottom). 
The weights are computed as the mean across ten utterances per emotion and their standard deviations are illustrated with error bars.  Positive component weights account for a prediction of the target emotion.
In contrast, negatively weighted components lead the model to predict a different emotion.
\vspace{-6pt}}
\label{fig:EmoLIME}
\vspace{-0.2cm}
\end{figure*}
LIME explains the predictions of any classiﬁer or regressor by treating it as a black-box and approximating it locally with an interpretable model \cite{ribeiro_2016}.
Explanations are generated by perturbing the input and training a surrogate model that assigns weights to each input component. 
Fig.~\ref{fig:schematic} depicts a functional block diagram of EmoLIME. 
The raw audio input is decomposed into frequency segments in the first step. 
We let $\mathbf{x}\in\mathbb{R}^d$ denote the original input representation, and $\mathbf{x'}\in\{0, 1\}^{d'}$ denotes a binary vector for its spectral decomposition indicating the presence or absence of the individual input components \cite{ribeiro_2016}. 
Training data for the surrogate model is generated by perturbing the input audio by randomly setting entries in $\mathbf{x'}$ to zero, resulting in $n$ training samples $\mathbf{z'} \in \{0, 1\}^{d'}$.
The loss function of the surrogate model, $g$, is a locally weighted square loss, given by:
\begin{equation}\label{eq:lime_loss}
    L(f,g,\pi_x) = \sum_{\mathbf{z}, \mathbf{z'} \in \mathbb{Z}} \pi_x (\mathbf{z}) (f(\mathbf{z}) - g(\mathbf{z'}))^2, 
\end{equation}
where $f$ is the black-box model and $\pi_x(\mathbf{z})$ is an exponential kernel learned over cosine distance, which accounts for the distance between the perturbed training samples $\mathbf{z}$ and the original input $\mathbf{x}$.
Hence, input samples $\mathbf{z}$ get predictions using $f$, and we weigh them by the proximity to the input being explained.
The implementation of EmoLIME builds on 
CoughLIME\footnote{\url{https://github.com/glam-imperial/CoughLIME}} \cite{wullenweber2022coughlime} and the LIME Python module \cite{Brinch_2010}, and it requires the prediction function to output logits rather than class labels.
To accommodate multiple classes, separate prediction functions were defined for each class to perform binary classification and output the class probability.
The surrogate model is obtained using Ridge regression as is the default in LimeBase\footnote{\url{https://github.com/marcotcr/lime}}.

To investigate hand-crafted vs.~deep features, two models are included in the analysis; a linear support vector classifier (SVC) trained on ComParE \cite{schuller_2016_ComParE} features and one trained on embeddings extracted from the last hidden states of a pre-trained Wav2Vec 2.0 (wav2vec2) model \cite{wagner2022model}, referred to as \texttt{ComParE-SVC} and \texttt{wav2vec2-SVC}, respectively.
Both models are trained on features using Leave-One-Speaker-Out (LOSO) cross-validation on the subsection of the datasets containing the emotions: happiness, anger, sadness, and neutral.
Hence, six separate models are trained; one for each combination of the two features and three datasets.
The models correctly classified the utterances included in the analysis, hence, the \textit{positive} class is the correct emotion while the \textit{negative} class consists of any other emotion.
This reasoning aligns well with the One-vs-Rest classification strategy that splits a multi-class classification into one binary classification problem per class.

\begin{table*}[!htb]
\centering
\fontsize{7.5pt}{9.5pt}\selectfont
\begin{tabular}{|cccc|c|cccc|c|cccc|}
\cline{1-4} \cline{6-9} \cline{11-14}
\multicolumn{4}{|c|}{\textbf{\begin{tabular}[c]{@{}c@{}}Two-sample Cramer test\\ ComPare vs. wav2vec2\end{tabular}}}                                                                              &  & \multicolumn{4}{c|}{\textbf{\begin{tabular}[c]{@{}c@{}}Two-sample Cramer test\\ ComPare: Dataset 1 vs. Dataset 2\end{tabular}}}                                                              &  & \multicolumn{4}{c|}{\textbf{\begin{tabular}[c]{@{}c@{}}Two-sample Cramer test\\ wav2vec2: Dataset 1 vs. Dataset 2\end{tabular}}}                                                                 \\ \cline{1-4} \cline{6-9} \cline{11-14} 
\multicolumn{1}{|c|}{\textbf{\begin{tabular}[c]{@{}c@{}}Dataset /\\ Emotion\end{tabular}}} & \multicolumn{1}{c|}{\textbf{Statistic}} & \multicolumn{1}{c|}{\textbf{Crit. Val.}} & \textbf{P-val.} &  & \multicolumn{1}{c|}{\textbf{\begin{tabular}[c]{@{}c@{}}Datasets /\\ Emotion\end{tabular}}} & \multicolumn{1}{c|}{\textbf{Statistic}} & \multicolumn{1}{c|}{\textbf{Crit. Val.}} & \textbf{P-val.} &  & \multicolumn{1}{c|}{\textbf{\begin{tabular}[c]{@{}c@{}}Datasets /\\ Emotion\end{tabular}}} & \multicolumn{1}{c|}{\textbf{Statistic}} & \multicolumn{1}{c|}{\textbf{Crit. Val.}} & \textbf{P-val.} \\ \cline{1-4} \cline{6-9} \cline{11-14} 
\multicolumn{1}{|c|}{EDB /}                                                                & \multicolumn{1}{c|}{}               & \multicolumn{1}{c|}{}                    &                 &  & \multicolumn{1}{c|}{EDB vs. RV /}                                                         & \multicolumn{1}{c|}{}               & \multicolumn{1}{c|}{}                    &                 &  & \multicolumn{1}{c|}{EDB vs. RV /}                                                         & \multicolumn{1}{c|}{}               & \multicolumn{1}{c|}{}                    &                 \\
\multicolumn{1}{|c|}{A}                                                                    & \multicolumn{1}{c|}{0.88}           & \multicolumn{1}{c|}{0.44}                & \textbf{\textbf{$<$0.01*}}        &  & \multicolumn{1}{c|}{A}                                                                    & \multicolumn{1}{c|}{0.32}           & \multicolumn{1}{c|}{0.31}                & $<$0.05*        &  & \multicolumn{1}{c|}{A}                                                                    & \multicolumn{1}{c|}{0.52}           & \multicolumn{1}{c|}{0.49}                & \textbf{$<$0.05*}        \\
\multicolumn{1}{|c|}{H}                                                                    & \multicolumn{1}{c|}{0.32}           & \multicolumn{1}{c|}{0.56}                & 0.32            &  & \multicolumn{1}{c|}{H}                                                                    & \multicolumn{1}{c|}{0.33}           & \multicolumn{1}{c|}{0.40}                & 0.10            &  & \multicolumn{1}{c|}{H}                                                                    & \multicolumn{1}{c|}{0.64}           & \multicolumn{1}{c|}{0.48}                & \textbf{$<$0.05*}        \\
\multicolumn{1}{|c|}{S}                                                                    & \multicolumn{1}{c|}{0.29}           & \multicolumn{1}{c|}{0.21}                & \textbf{$<$0.05*}        &  & \multicolumn{1}{c|}{S}                                                                    & \multicolumn{1}{c|}{0.88}           & \multicolumn{1}{c|}{0.39}                & \textbf{$<$0.01*}        &  & \multicolumn{1}{c|}{S}                                                                    & \multicolumn{1}{c|}{0.22}           & \multicolumn{1}{c|}{0.29}                & 0.17            \\
\multicolumn{1}{|c|}{N}                                                                    & \multicolumn{1}{c|}{0.45}           & \multicolumn{1}{c|}{0.36}                & \textbf{$<$0.05*}        &  & \multicolumn{1}{c|}{N}                                                                    & \multicolumn{1}{c|}{0.80}           & \multicolumn{1}{c|}{0.42}                & \textbf{$<$0.01*}        &  & \multicolumn{1}{c|}{N}                                                                    & \multicolumn{1}{c|}{1.32}           & \multicolumn{1}{c|}{0.56}                & \textbf{$<$0.01*}        \\ \cline{1-4} \cline{6-9} \cline{11-14} 
\multicolumn{1}{|c|}{RV /}                                                                 & \multicolumn{1}{c|}{}               & \multicolumn{1}{c|}{}                    &                 &  & \multicolumn{1}{c|}{EDB vs. IE /}                                                         & \multicolumn{1}{c|}{}               & \multicolumn{1}{c|}{}                    &                 &  & \multicolumn{1}{c|}{EDB vs. IE /}                                                         & \multicolumn{1}{c|}{}               & \multicolumn{1}{c|}{}                    &                 \\
\multicolumn{1}{|c|}{A}                                                                    & \multicolumn{1}{c|}{0.25}           & \multicolumn{1}{c|}{0.37}                & 0.22            &  & \multicolumn{1}{c|}{A}                                                                    & \multicolumn{1}{c|}{0.22}           & \multicolumn{1}{c|}{0.14}                & \textbf{$<$0.01*}        &  & \multicolumn{1}{c|}{A}                                                                    & \multicolumn{1}{c|}{0.74}           & \multicolumn{1}{c|}{0.43}                & \textbf{$<$0.01*}        \\
\multicolumn{1}{|c|}{H}                                                                    & \multicolumn{1}{c|}{0.15}           & \multicolumn{1}{c|}{0.25}                & 0.35            &  & \multicolumn{1}{c|}{H}                                                                    & \multicolumn{1}{c|}{0.51}           & \multicolumn{1}{c|}{0.58}                & 0.07            &  & \multicolumn{1}{c|}{H}                                                                    & \multicolumn{1}{c|}{0.34}           & \multicolumn{1}{c|}{0.54}                & 0.23            \\
\multicolumn{1}{|c|}{S}                                                                    & \multicolumn{1}{c|}{0.21}           & \multicolumn{1}{c|}{0.38}                & 0.39            &  & \multicolumn{1}{c|}{S}                                                                    & \multicolumn{1}{c|}{0.70}           & \multicolumn{1}{c|}{0.37}                & \textbf{$<$0.01*}        &  & \multicolumn{1}{c|}{S}                                                                    & \multicolumn{1}{c|}{0.18}           & \multicolumn{1}{c|}{0.25}                & 0.17            \\
\multicolumn{1}{|c|}{N}                                                                    & \multicolumn{1}{c|}{0.14}           & \multicolumn{1}{c|}{0.44}                & 0.65            &  & \multicolumn{1}{c|}{N}                                                                    & \multicolumn{1}{c|}{0.40}           & \multicolumn{1}{c|}{0.35}                & \textbf{$<$0.05*}        &  & \multicolumn{1}{c|}{N}                                                                    & \multicolumn{1}{c|}{0.87}           & \multicolumn{1}{c|}{0.41}                & \textbf{$<$0.01*}        \\ \cline{1-4} \cline{6-9} \cline{11-14} 
\multicolumn{1}{|c|}{IE /}                                                                 & \multicolumn{1}{c|}{}               & \multicolumn{1}{c|}{}                    &                 &  & \multicolumn{1}{c|}{RV vs. IE /}                                                          & \multicolumn{1}{c|}{}               & \multicolumn{1}{c|}{}                    &                 &  & \multicolumn{1}{c|}{RV vs. IE /}                                                          & \multicolumn{1}{c|}{}               & \multicolumn{1}{c|}{}                    &                 \\
\multicolumn{1}{|c|}{A}                                                                    & \multicolumn{1}{c|}{0.77}           & \multicolumn{1}{c|}{0.24}                & \textbf{$<$0.01*}        &  & \multicolumn{1}{c|}{A}                                                                    & \multicolumn{1}{c|}{0.60}           & \multicolumn{1}{c|}{0.32}                & \textbf{$<$0.01*}        &  & \multicolumn{1}{c|}{A}                                                                    & \multicolumn{1}{c|}{0.22}           & \multicolumn{1}{c|}{0.25}                & 0.08            \\
\multicolumn{1}{|c|}{H}                                                                    & \multicolumn{1}{c|}{0.24}           & \multicolumn{1}{c|}{0.49}                & 0.33            &  & \multicolumn{1}{c|}{H}                                                                    & \multicolumn{1}{c|}{0.95}           & \multicolumn{1}{c|}{0.44}                & \textbf{$<$0.01*}        &  & \multicolumn{1}{c|}{H}                                                                    & \multicolumn{1}{c|}{0.68}           & \multicolumn{1}{c|}{0.45}                & \textbf{$<$0.01*}        \\
\multicolumn{1}{|c|}{S}                                                                    & \multicolumn{1}{c|}{0.26}           & \multicolumn{1}{c|}{0.35}                & 0.16            &  & \multicolumn{1}{c|}{S}                                                                    & \multicolumn{1}{c|}{0.44}           & \multicolumn{1}{c|}{0.46}                & 0.06            &  & \multicolumn{1}{c|}{S}                                                                    & \multicolumn{1}{c|}{0.19}           & \multicolumn{1}{c|}{0.30}                & 0.24            \\
\multicolumn{1}{|c|}{N}                                                                    & \multicolumn{1}{c|}{0.25}           & \multicolumn{1}{c|}{0.35}                & 0.17            &  & \multicolumn{1}{c|}{N}                                                                    & \multicolumn{1}{c|}{0.30}           & \multicolumn{1}{c|}{0.46}                & 0.14            &  & \multicolumn{1}{c|}{N}                                                                    & \multicolumn{1}{c|}{0.41}           & \multicolumn{1}{c|}{0.38}                & \textbf{$<$0.05*}        \\ \cline{1-4} \cline{6-9} \cline{11-14} 
\end{tabular}
\vspace{3pt}
\caption{Two-sample Multivariate Nonparametric Cramer-Test. EDB: EMODB, RV: RAVDESS, IE: IEMOCAP, A. Anger, H: Happiness, S: Sadness, N: Neutral. Tests where the null hypothesis is rejected are marked by *.}
\label{tab:Cramer_test}
\vspace{-0.8cm}
\end{table*}

\section{Experimental resources}
EmoLIME explanations were generated for ten randomly selected utterances per emotion balanced across speakers in the datasets, as visualized in Fig. \ref{fig:EmoLIME}.
The random seed is kept constant to ensure the input data is perturbed similarly when comparing the models.
We used the following datasets in this work: 
\begin{enumerate*}
\item {\bf EMODB} (Berlin Database of Emotional Speech) \cite{Burkhardt_2005_EMODB} contains acted emotional speech in German. 
Ten speakers (ﬁve male and ﬁve female) participated in the study each producing ten utterances that were a mix of short and longer sentences. 
In total, the database contains 535 recordings. 
\item {\bf RAVDESS} (Ryerson Audio-Visual Database of Emotional Speech and Song) \cite{livingstone_2018_Ravdess} is an audio-visual database containing enacted emotional speech and song from 24 professional actors (12 female and 12 male). 
The corpus contains 7356 recordings in English with a neutral North American accent.
\item {\bf IEMOCAP} (The Interactive Emotional Dyadic Motion Capture) \cite{Busso_2008_IEMOCAP} database is an acted, multimodal database in English. 
Ten actors (five male and five female) perform improvisations or scripted scenarios, specifically selected to elicit emotional expressions.
The database includes 1277 recorded utterances.
\end{enumerate*}

\section{Results and Discussion}


The spectral decomposition segment the audio into eight equally sized spectral components in the frequency range between 0 to 8 kHz.
Only true predictions are included in the visualizations, hence positive weights correspond to components that yield the model towards predicting the true class. 
Intuitively, low-pitch speech can be associated with low valence emotions, such as anger and sadness. 
In contrast, high pitch is usually associated with high valence emotions, such as happiness. 
For EMODB, this was indeed the observation for the model trained on deep features, but not for the model built on hand-crafted features as exemplified in Figs.~\ref{fig:spectral_decomps_1} and \ref{fig:spectral_decomps_2}. 
 

We quantify the average spectral decomposition weights across a selection EMODB, RAVDESS, and IEMOCAP of utterances in Fig.~\ref{fig:EmoLIME}.
Although, the fundamental frequency of the human voice lies in the range of 90 to 155 Hz for men and between 165 to 255 Hz for women,  research has shown that high-frequency components up to and above 7 kHz play a role in human hearing and perception \cite{jacewicz2023a}.
Very high-pitch components do not contribute significantly and are assigned close-to-zero weights by the EmoLIME algorithm. 
Some key takeaways from Figure \ref{fig:EmoLIME} are:
\begin{enumerate*}[label=(\roman*)]
    \item {\it Deep features}: Low-pitch components ($<$3 kHz) contribute most to predicting angry and sad emotions. 
    \item {\it Deep features}: High-pitch components ($<$3 kHz) tend to account for more in the prediction of high-arousal emotions (happy, angry) compared to low-arousal emotions (sad, neutral).
    \item {\it Hand-crafted features}: Spectral weights for very high-pitch components ($<$4 kHz) are closer to zero when compared to the deep features model, except for sad emotions.
    \item {\it General trend}: Indications that the EmoLIME technique is more robust across models than across datasets for the same emotion.
\end{enumerate*}

To {\it statistically} test observation (iv) above, we perform a non-parametric Cramer-Test \cite{baringhaus2004a} for the multivariate two-sample problem with the {\it null hypothesis: the two samples come from the same underlying distribution} at $\alpha = 0.05$ significance level.
The spectral weight distributions consist of 10 samples and 8 dimensions per emotion, and results are listed in Table \ref{tab:Cramer_test}.
The null hypothesis is accepted in 8/12 (67\%) possible tests for the same dataset but different models. 
In comparison, the null hypothesis is accepted in 9/24 (38\%) possible tests for the same model but different datasets. This further reinforces our observation that EmoLIME is less robust to distribution shifts.

\section{Conclusion} 
Expressing and interpreting emotions is a highly subjective process, and investigating XAI methods for SER forces us to reflect on how humans perceive emotions through speech.
It remains a substantial challenge to evaluate XAI techniques on more complex speech tasks owing to the involvement of multiple components within the model such as general language models, the challenge of reliably mapping from speech input to objective ground truths, and the variability due to speakers, language, culture, etc., which is unique to speech signals. 
We propose EmoLIME, a LIME based XAI method for SER models, and demonstrate that the method can produce explanations that are well-aligned with human intuition.
Using EmoLIME, an exploration of average spectral decomposition weights for models based on hand-crafted and deep features was undertaken.
The emotional representations learned by the pre-trained model align well with the intuitive connection between pitch and high vs.~low valence emotions.
To further the development of XAI techniques for SER towards a more comprehensive understanding of model predictions, one could consider incorporating global explanations through gradient-based techniques or SHAP, in addition to the local explanations in EmoLIME.

\section*{Acknowledgment}
Partially funded by the European Union grants under the Marie Skłodowska-Curie Grant Agreements Nos: 101169474 (AlignAI) and 101081465 (AUFRANDE). Views and opinions expressed are however those of the author(s) only and do not necessarily reflect those of the European Union or the Research Executive Agency. Neither the European Union nor the Research Executive Agency can be held responsible for them.

\bibliographystyle{IEEEbib}
\bibliography{refs}

\end{document}